%% file: main.tex
\documentclass[a4paper, amsfonts, amssymb, amsmath, preprint, showkeys, nofootinbib, twoside]{revtex4-1}
\usepackage[english]{babel}
\usepackage[utf8]{inputenc}
\usepackage[colorinlistoftodos, color=green!40, prependcaption]{todonotes}
\input{preamble}

\begin{document}
\title{Five-dimensional non-Abelian supersymmetric Chern-Simons action in Projective Superspace}

\author{Ariunzul Davgadorj}
    \email{ariunzul.d@gmail.com}

\affiliation{Department of Mathematics and Physics, Technical University of Kielce, al.Tysi\k aclecia Pa\'nstwo Polskiego 7, 25-314 Kielce, Poland}

\date{\today}

\begin{abstract}

In this paper, we derive five-dimensional non-Abelian $\mathcal{N}=1$ Chern-Simons action from its proposed definition of variation with respect to an infinitesimal deformation of the vector prepotential in a projective superspace setting. It has been a long-standing open problem how to integrate this variation. Now, thanks to the simplicity of a projective superspace technique, we are finally able to obtain this term.  Our final result is expressed in terms of the real prepotential multiplets and projective covariant derivatives. The action reduces to the correct formula in the Abelian limit. As a bonus, we also give the supersymmetric Yang-Mills action with a half measure.
\end{abstract}

\vspace{3cm}

\maketitle
\tableofcontents

\input{sections/Intro.tex}

\input{sections/section1.tex} 
\input{sections/section2.tex}

\input{Acknowledgements.tex}

\appendix

\input{sections/Appendix.tex}

\bibliography{Bibliography}{}
\bibliographystyle{unsrtnat}

\end{document}

%% file: preamble.tex
\usepackage{amsthm}
\usepackage{mathtools}
\usepackage{physics}
\usepackage{xcolor}
\usepackage{graphicx}
\usepackage[left=23mm,right=13mm,top=35mm,columnsep=15pt]{geometry} 
\usepackage{hyperref}
\usepackage{adjustbox}
\usepackage{placeins}
\usepackage[T1]{fontenc}
\usepackage{lipsum}
\usepackage{csquotes}
\usepackage{tikz}
\newcommand{\overbar}[1]{\mkern 1.5mu\overline{\mkern-1.5mu#1\mkern-1.5mu}\mkern 1.5mu}
\renewcommand{\bar}{\overbar}
\renewcommand{\tilde}{\widetilde}

\newcommand{\nn}{\nonumber}


%% file: sections/Intro.tex
\section{Introduction}\label{sec:Intro}
The most recognizable prototypical topological field theory is the Chern-Simons theory in 3-dimensions . This theory is solvable exactly and Witten showed that the expectation value of the topologically invariant observables, e.g., Wilson loops, is equivalent to the Jones polynomial of knot homology invariants emerging from the Chern-Simons theory on any 3-manifold \cite{Witten:1988hf}. Uplifting this approach (describing knot invariants via Chern-Simons theory on the boundary of a higher-dimensional manifold) to five dimensions has also been done, leading to the categorization of the Jones polynomial \cite{Khovanov:1999qla, Witten:2010cx, Gaiotto:2011nm, Witten:2011xiq}.

Five-dimensional supersymmetric gauge theories have been discussed in different contexts in the past, such as superconformal fixed points as an infinite coupling limit of super Yang-Mills theory (SYM) with various gauge groups, particularly with $U(N)$, $SU(N)$ groups, which allow the Chern-Simons term in the action \cite{Seiberg:1996bd, Intriligator:1997pq, Minahan:2013jwa, Minahan:2014hwa}; or due to their relation to the six-dimensional $\mathcal{N}=(2,0)$ superconformal field theories. In this case, the UV completion of the 5d effective field theory is a 6d theory, whose one dimension is compactified on a circle \cite{Witten:1995ex, Douglas:2010iu, Lambert:2012qy}. It was also argued that the Chern-Simons term can be obtained from the super Yang-Mills theory with massive hypermultiplets by integrating out the hypermultiplet \cite{Seiberg:1996bd, Intriligator:1997pq}. 

In harmonic superspace settings \cite{Galperin:1984av, Galperin:2001seg}, one-loop effective action for a hypermultiplet with a background vector multiplet has been obtained, and as a result the Chern-Simons action at the quantum level has been generated. \cite{Kuzenko:2006ek, Buchbinder:2015wpa}. The leading order of this quantum correction agrees with the action formula proposed in \cite{Zupnik:1999iy}. In that paper Zupnik defined the variation of the non-Abelian Chern-Simons action with respect to the infinitesimal prepotential deformation. However, the integrated form of this action has not been known. The purpose of this paper is to finally find an answer to this question. In the Abelian case, the action was obtained in \cite{Zupnik:1999iy, Kuzenko:2005sz}. The superform approach to the non-Abelian case was introduced in \cite{Kuzenko:2013rna}.

In this paper, we will employ the projective superspace \cite{Karlhede:1984vr} approach to this long-standing problem. Projective superspace formalism is a manifestly supersymmetric method developed for extended supersymmetric theories with an $R$-symmetry group containing one or more factors of $SU(2)$. In the case of four-dimensions, the ordinary superspace $\mathbb{R}^{4|8}$ is appended by an auxiliary $\mathbb{CP}^1=\frac{SU(2)}{U(1)}$ factor, on which all the fields depend holomorphically. Contour integrals are then used on this extra coordinate to construct invariant actions. A comparison between Projective and Harmonic superspace can be found in \cite{Kuzenko:1998xm, Butter:2012ta}. A hybrid formalism between them was also developed in \cite{Jain:2009aj, Jain:2010gm, Jain:2012jx, Jain:2013hua}. Applications of the Projective superspace method, for instance, to find hypermultiplet contributions to the effective action \cite{Gonzalez-Rey:1997msl, Gonzalez-Rey:1997pxs} or in the theory of Abelian super Yang-Mills \cite{Lindstrom:1989ne}, in the quantization of the Abelian vector multiplet \cite{Gonzalez-Rey:1997spr} and in the non-Abelian super Yang-Mills theory \cite{Davgadorj:2017ezp} can be found. New techniques to treat, more precisely, how to separate the field $e^{V(\zeta)}$ (constructed from the unconstrained gauge prepotential $V(\zeta)$) into negative and positive powers of $\mathbb{CP}^1$ holomorphic coordinate $\zeta$ was developed in \cite{Davgadorj:2023lfv}. Considering that all objects need to be expanded in powers of this object $X=e^V-1$ at different $\zeta$-coordinates, particular separation/projection methods make a difference, leading to significant simplification of expressions for the field strength and Lagrangians. One of the hallmarks of Projective superspace is its straightforwardness in reducing the higher supersymmetric expression into lower supersymmetric components, the new method enables this reduction efficiently. The power of this technique was tested and applied to the $\mathcal{N}=3$ and $\mathcal{N}=4$ gauge theories in three-dimensions as well as in four-dimensional, 8-supercharge Yang-Mills action in that paper. 

In this article, in section III, we will again utilize the power of this non-symmetric separation technique of \cite{Davgadorj:2023lfv} in deriving the five-dimensional $\mathcal{N}=1$ non-Abelian Chern-Simons term from its well-defined variation formula. But before that, for the sake of completeness, in Section II, we will describe the Yang-Mills action with a projective half-measure. This formula will be proven to be correct, when we reduce it to the 4-dimensions. In order to be self-contained, we also include the non-symmetric splitting techniques for the prepotential multiplet $e^V$ and subsequently for the field strength in the appendices.

%% file: sections/section1.tex
\section{Yang-Mills action with a half measure}\label{sec:YM}
\subsection{5-dimensional superalgebra}
Let us start with a comment on spinors in 5-dimensions. Here, one cannot have Majorana spinors like in $d=2,3,4$, because the reality condition that is $\psi=C(\Gamma^0)^T\psi^*$ has no solution. Therefore, one constructs pairs of Dirac spinors that transform to each other under charge conjugation: $\psi_1=C(\Gamma^0)^T\psi^*_2$ and $\psi_2=-C(\Gamma^0)^T\psi^*_1$. Concisely: $\psi_i=\epsilon^{ij}C(\Gamma^0)^T\psi^*_j$ This is called the symplectic Majorana spinors / condition.  
The antisymmetric 5D $\Gamma$ matrices that satisfy a relation $C\Gamma^a=-(C\Gamma^a)^T$ are:
\begin{align}
    &C\Gamma^0=\begin{pmatrix}
    0 & \begin{pmatrix}
           0 & 1\\
           -1 & 0
        \end{pmatrix} \\
    \begin{pmatrix}
       0 & 1\\
       -1 & 0
    \end{pmatrix} & 0 
    \end{pmatrix}\;\;;\;\;C\Gamma^1=\begin{pmatrix}
    0 & \begin{pmatrix}
           -1 & 0\\
           0 & 1
        \end{pmatrix} \\
    \begin{pmatrix}
       1 & 0\\
       0 & -1
    \end{pmatrix} & 0 
    \end{pmatrix}\;\;;\nn \\
    &C\Gamma^2=\begin{pmatrix}
    0 & \begin{pmatrix}
           i & 0\\
           0 & i
        \end{pmatrix} \\
    \begin{pmatrix}
       -i & 0\\
       0 & -i
    \end{pmatrix} & 0 
    \end{pmatrix}\;\;;\;\;C\Gamma^3=\begin{pmatrix}
    0 & \begin{pmatrix}
           0 & -1\\
           -1 & 0
        \end{pmatrix} \\
    \begin{pmatrix}
       0 & 1\\
       1 & 0
    \end{pmatrix} & 0 
    \end{pmatrix}\;\;;\nn \\
    &C\Gamma^5=\begin{pmatrix}
        \begin{pmatrix}
           0 & -i\\
           i & 0
        \end{pmatrix} & 0 \\
        0 & \begin{pmatrix}
              0 & i\\
              -i & 0
            \end{pmatrix} 
    \end{pmatrix}\;=\;\begin{pmatrix}
                          -i\epsilon^{\alpha\beta} & 0 \\
                          0 & i\epsilon^{\dot{\alpha}\dot{\beta}}
                      \end{pmatrix}
\end{align}
Here the antisymmetric $C=\epsilon^{\hat{\alpha}\hat{\beta}}=\begin{pmatrix}
\epsilon^{\alpha\beta} & 0 \\
0 & \epsilon^{\dot{\alpha}\dot{\beta}}
\end{pmatrix}$ and the inverse $C^{-1}=\epsilon_{\hat{\alpha}\hat{\beta}}$ serve as charge conjugation matrices. $\epsilon$ matrices are used to raise and lower indices. $\epsilon^{\hat{\alpha}\hat{\beta}}\epsilon_{\hat{\alpha}\hat{\beta}}=-4$.\\

Superfields in the superspace $\mathbb{R}^{5|8}$ depend on $(x^{\hat{a}},\theta^{\hat{\alpha}}_i)$ coordinates, where $\hat{a}$ and $\hat{\alpha}$ are the 5-vector and 4-spinor indices of the Lorentz group $SO(4,1)$, while $i$ is the index of the $SU(2)$ automorphism group. The $SU(2)_R$ index reflects an internal symmetry and distinguishes components within the doublet. This internal structure allows us to organize the superspace spinorial derivatives as $D^i_{\hat{\alpha}}$ respecting the $SU(2)_R$ symmetry. The superalgebra for the ungauged derivatives is:
\begin{equation}
    \{D^i_{\hat{\alpha}},D^j_{\hat{\beta}}\}=i\epsilon^{ij}\nabla_{\hat{\alpha}\hat{\beta}}\;,\;\textit{where}\;\;\nabla_{\hat{\alpha}\hat{\beta}}=(C\Gamma^{\hat{a}})_{\hat{\alpha}\hat{\beta}}\partial_{\hat{a}}
\end{equation}

While the gauged covariant derivatives ($\mathcal{D}^i_{\hat{\alpha}}=D^i_{\hat{\alpha}}+\Gamma^i_{\hat{\alpha}}\;,\;\Gamma^i_{\hat{\alpha}}$ is the spinorial gauge connection) satisfy the following algebra:
\begin{align}
    \{\mathcal{D}^i_{\hat{\alpha}},\mathcal{D}^j_{\hat{\beta}}\}=i\epsilon^{ij}\left(\boldsymbol{\nabla}_{\hat{\alpha}\hat{\beta}}+\epsilon_{\hat{\alpha}\hat{\beta}}\mathbb{W}\right)\;,\;\textit{where}\;\;\boldsymbol{\nabla}_{\hat{\alpha}\hat{\beta}}=(C\Gamma^{\hat{a}})_{\hat{\alpha}\hat{\beta}}{\mathcal{D}}_{\hat{a}}
\end{align}
Above, $\partial_{\hat{a}}$ and $\mathcal{D}_{\hat{a}}$ denote respectively ungauged and gauged spacetime derivatives. The bold symbols also represent gauged operators. $\mathbb{W}$ is the field strength in vector representation.\\

Now, we define a projective subspace, where our constrained superfields live. This can be achieved by introducing a coordinate on an auxiliary $\mathbb{CP}^1$ manifold and parameterizing fields and covariant derivatives with this coordinate. The operator that constrains projective superfields is:
\begin{equation}
\nabla_{\hat{\alpha}}\equiv D^1_{\hat{\alpha}}+\zeta D^2_{\hat{\alpha}}
\end{equation}
The complex coordinate $\zeta$ is a part of an isospinor $\boldsymbol{u}$.
\begin{equation}\label{isospinor}
\boldsymbol{u}^T=\frac{1}{u^1}(u^1\;,\;u^2) \;\;\textit{and}\;\; \zeta\equiv\frac{u^2}{u^1}
\end{equation}
Then, the nabla operator is written as the product of the projective isospinor and the spinorial covariant derivatives. \(\nabla_{\hat{\alpha}}=\boldsymbol{u}^TD_{i{\hat{\alpha}}}\). Therefore, $\nabla_{\hat{\alpha}}$ transforms covariantly under $SU(2)_R$ with $1/u^1$ scaling. From the superalgebra, it is clear that the ungauged spinorial $\nabla$ operators anticommute.
\begin{equation}
 \{\nabla_{\hat{\alpha}}(\zeta),\nabla_{\hat{\beta}}(\zeta)\}=0 
\end{equation}
The anticommutation rule also applies to the gauge covariantized operator: $\boldsymbol{\nabla}_{\hat{\alpha}}\equiv \mathcal{D}_{1\hat{\alpha}}+\zeta \mathcal{D}_{2\hat{\alpha}}$:
\begin{align}
 \{\boldsymbol\nabla_{\hat{\alpha}}(\zeta),\boldsymbol\nabla_{\hat{\beta}}(\zeta)\}=0 
\end{align}
This does not tell us much except that this constraint can be explained as the integrability condition for the existence of the projective superfields $\mathcal{P}$, which are defined to be annihilated by the $\nabla$ operators: $\nabla_{\hat{\alpha}}\mathcal{P}=0$, and then the gauged nablas can be related with the ungauged ones through: $\boldsymbol{\nabla}_{\hat{\alpha}}=\nabla_{\hat{\alpha}}+\Gamma_{\hat{\alpha}}(\zeta)=e^U\nabla_{\hat{\alpha}} e^{-U}=e^{-\bar{U}}\nabla_{\hat{\alpha}} e^{\bar{U}}$. Here, $e^V=e^{\bar{U}}e^U$. $V$ is our prepotential superfield, which in projective superspace we call a tropical multiplet, meaning its $\zeta$-dependance is singular at both north and south pole of the $\mathbb{CP}^1$ manifold: \(V(\zeta)=\sum_{n=-\infty}^\infty v_n\zeta^n\). $V$ and subsequently $e^V$ are projective superfields, $e^U$ contains only positive powers of $\zeta$ components plus the half of constant terms, hence arctic; $e^{\bar{U}}$ contains only negative powers of $\zeta$ plus the half of constant terms, therefore antarctic multiplet, but $e^U$ and $e^{\bar{U}}$ are not annihilated by the nabla operator, so they are not projective superfields. \\

To produce field strengths, we now take the $\zeta$'s at different points.
\begin{align}
   \{\boldsymbol\nabla_{\hat{\alpha}}(\zeta_1),\boldsymbol\nabla_{\hat{\beta}}(\zeta_2)\}=i(\zeta_1-\zeta_2)\left(\boldsymbol{\nabla}_{\hat{\alpha}\hat{\beta}}+\epsilon_{\hat{\alpha}\hat{\beta}}\mathbb{W}\right)
\end{align}
We do the same trick as we did in 4D, take a $\zeta$-derivative with respect to one of the $\zeta$'s and act with $e^{U}$ and $e^{-U}$ from both sides. However, the difference from the 4-dimensional case is that now we have an additional term $\nabla_{\hat{\alpha}\hat{\beta}}$, but this will not produce anything problematic, since one can see that $\epsilon^{\hat{\alpha}\hat{\beta}}(C\Gamma^{\hat{a}})_{\hat{\alpha}\hat{\beta}}=0$, therefore:
\begin{align}
\epsilon^{\hat{\alpha}\hat{\beta}}\;\;\vert\;\;\{\nabla_{\hat{\alpha}},[e^{-U}\partial_\zeta e^U,\nabla_{\hat{\beta}}]\}&=ie^{-U}(C\Gamma^{\hat{a}})_{\hat{\alpha}\hat{\beta}}\partial_{\hat{a}}e^U+i\epsilon_{\hat{\alpha}\hat{\beta}}e^{-U}\mathbb{W}e^U \nonumber \\
-4i\mathcal{W}&=-\epsilon^{\hat{\alpha}\hat{\beta}}\nabla_{\hat{\alpha}}\nabla_{\hat{\beta}}A_\zeta
\end{align}
where, as usual, we define the arctic field strength $\mathcal{W}=e^{-U}\mathbb{W}e^U$ (the field strength in vector representation has no $\zeta$-dependence, while in the arctic representation, contains all positive powers of $\zeta$-expansion plus the constant term) and the covariant $\zeta$-connection $A_\zeta=e^{-U}(\partial_\zeta e^U)$ ($\mathcal{D}_\zeta=\partial_\zeta+A_\zeta=e^{-U}\partial_\zeta e^U$). Then the above relation can be written compactly:
\begin{equation}
    \mathcal{W}=\frac{1}{2}i\nabla^2 A_\zeta
\end{equation}
Explicit derivation of the connection $A_\zeta$ for 4D can be found in \cite{Davgadorj:2017ezp}. In 5D, it is exactly the same.
\begin{align}
  &A_\zeta(\zeta_0)=\sum_{n=1}^\infty (-1)^{n+1} \oint d\zeta_1 \ldots \oint d\zeta_n \frac{X_1\ldots X_n}
{\zeta_{10}\zeta_{21}\ldots\zeta_{n,n-1}\zeta_{n0}}\;, \nonumber\\
&\textit{where} \;\; X_i=e^{V(\zeta_i)}-1\;;\;\; \frac{1}{\zeta_{ab}}=\frac{1}{\zeta_a}\sum_{n=0}^\infty \left(\frac{\zeta_b}{\zeta_a}\right)^n
\end{align}
On the other hand, if we acted with $e^{\bar{U}}$ and $e^{-\bar{U}}$ from both sides of the anticommutator above, we would have defined the antarctic covariant $\zeta$ connection $\tilde{A}_\zeta$ as: $\tilde{\mathcal{D}}_\zeta=\partial_\zeta+\tilde{A}_\zeta=e^{\bar U}\partial_\zeta e^{-\bar U}$. Accordingly, in this case we can find the antarctic field strength $\widetilde{\mathcal{W}}=e^{\bar{U}}\mathbb{W}e^{-\bar{U}}$, containing all negative powers of $\zeta$-expansion plus the constant term:
\begin{align}
    &\widetilde{\mathcal{W}}=\frac{1}{2}i\nabla^2 \tilde{A}_\zeta \;\;,\;\; \text{where the antarctic connection is:} \\
    &\tilde{A}_\zeta(\zeta_0)= \sum_{n=1}^\infty(-1)^{n+1} \oint d\zeta_1 \ldots \oint d\zeta_n \frac{X_1\ldots X_n}
{\zeta_{01}\zeta_{21}\ldots\zeta_{n,n-1}\zeta_{0n}}
\end{align}

Naturally, there is a relation between $A_\zeta$ and $\widetilde{A}_\zeta$ involving only $e^V$.
\begin{align}\label{useful}
&e^{-V}(\partial_\zeta e^V) = e^{-U}(\partial_\zeta e^{U}) + e^{-U}e^{-\bar U}(\partial_\zeta e^{\bar U}) e^{U} =
A_\zeta - e^{-V}\tilde{A}_\zeta e^{V}\;\ ,\nonumber\\
&e^V\mathcal{W}e^{-V}=\widetilde{\mathcal{W}}
\end{align}
\subsection{SYM action in terms of the field strength }
Zupnik has established long time ago in \cite{Zupnik:1999iy}, that the 8-supercharge super Yang-Mills theories with full measure have an universal form in terms of the prepotential. That is (in projective superspace description \cite{Jain:2012zx, Davgadorj:2017ezp}):
\begin{align}\label{5DfullYM}
    S_{YM} = \frac{1}{g_{YM}^2}\sum_{n=2}^{\infty} \frac{(-1)^n}{n}\tr\int d^5 x d^8\theta \oint d\zeta_1\ldots d\zeta_n
     \frac{X_1 X_2\ldots X_n}{\zeta_{21}\ldots \zeta_{n,n-1}\zeta_{1n}}
\end{align}
Then, one might want to construct the Yang-Mills action with a measure consisting of four supercovariant derivatives in terms of the field strength just like in four-dimensions. In 4D, one can construct a chiral/antichiral action with a chiral /antichiral half-measure. However, in five-dimensions a chiral description does not exist, as it is not Lorentz-covariant; therefore, a projective half-measure is an option. The full measure as projective covariant derivatives and as regular 5D, $\mathcal{N}=1$ covariant derivatives respectively: :
\(\int d^8\theta=\Delta^4\nabla^4=(D^{\hat{\alpha}}_1D_{1\hat{\alpha}})^2(D^{\hat{\alpha}}_2D_{2{\hat\alpha}})^2\)\;.\\
\(
\Delta_{\hat{\alpha}}\equiv D_{2{\hat\alpha}}-\frac{1}{\zeta} D_{1{\hat\alpha}}
\) is linearly independent from $\nabla_{\hat{\alpha}}$ and anticommute among themselves. $\nabla$ and $\Delta$ operators do not anticommute and we square them with the factor of 1/2: $\nabla^2=\frac{1}{2}\nabla^{\hat{\alpha}}\nabla_{\hat{\alpha}}$. Then, $\Delta^4$ will serve as the projective half-measure.\\

\noindent
We now need to construct a projective covariant superfield $\it{G}$ in terms of the field strength \cite{Kuzenko:2005sz}. 
\begin{equation}
    G=\nabla^4\{A_\zeta,\mathcal{W}\}\;\;,\;\;\text{where the \{\} brace indicates anticommutator} 
\end{equation}
The idea of applying Sohnius's prescription to construct a supersymmetric invariant action for an algebra with central charges \cite{Sohnius:1978fw} to a four-dimensional $\mathcal{N}=2$ harmonic superspace in the context of vector-tensor multiplet was developed in \cite{Dragon:1997za}. Later, this was applied  to describe five-dimensional super Yang-Mills action in \cite{Buchbinder:2015wpa}. 

In our case, we take a similar route; first, parametrizing the Grassmann coordinates with the $\mathbb{CP}^1$ parameter, such that our $\Delta$ operator that defines the projective measure is  proportional to the partial derivative with respect to this projective Grassmann variable.

Now, let us redefine Grassmann coordinates into projective ones.
\begin{align} \label{proj.theta}
    &\Theta_{\hat{\alpha}}=\theta_{\hat{\alpha}2}-\zeta\theta_{\hat{\alpha}1}\;\;\textit{such that}\;\;\Delta^{\hat{\alpha}}=D^{\hat{\alpha}}_2-\frac{1}{\zeta}D^{\hat{\alpha}}_1\;\sim\;\frac{\partial}{\partial\Theta_{\hat{
    \alpha}}}\\
    &\Theta^{\hat{\alpha}}=\theta^{\hat{\alpha}}_2-\zeta\theta^{\hat{\alpha}}_1\;\;\textit{such that}\;\;\Delta_{\hat{\alpha}}=D_{\hat{\alpha}2}-\frac{1}{\zeta}D_{\hat{\alpha}1}\;\sim\;\frac{\partial}{\partial\Theta^{\hat{
    \alpha}}}
\end{align}
Then, we define a projective, constant field $\tau$ that couples the projective covariant multiplet $G$, the coupling is gauge-invariant.
The constant field $\tau$ is defined by:
\begin{align}
    \Delta^2\tau\equiv\frac{1}{g_{YM}^2}\; \implies\; \tau=\frac{1}{g_{YM}^2}\Theta^2;\;\;\;\;\textit{also}\;\;\;\; \nabla\tau=0
\end{align}
$G_{YM}$ is a constrained non-Abelian YM multiplet: $\nabla G_{YM}=0$. The SYM action in the Sohnius prescription in projective language will become:
\begin{align}\label{YM5D}
    S_{YM}=\int d^5 x \oint \frac{d\zeta}{\zeta}\Delta^4\;\tau \;\rm{tr}\;G_{YM}=\frac{1}{g_{YM}^2}\rm{tr}\int d^5 x \oint \frac{d\zeta}{\zeta}\Delta^4 \Theta^2 G,\;\;\textit{where}\;\;G=\nabla^4\{A,\mathcal{W}\}
\end{align}
In the next part, we will show that this particular construction of action reduces to the expected expression in four-dimensional Yang-Mills action formula, proving that this is the correct form. 


\subsection{Reduction to 4-dimensions}
One could want to see how this above-received construction works out if we reduce the dimension, does it produce well-known 4-dimensional chiral actions? For this purpose, let us write down the 5-dimensional symplectic Majorana spinors as a pair of spinors that are constructed from 4-dimensional spinors. The pairs are related to each other with the antisymmetric tensor $\epsilon^{ij}$.
Spinor relations between $\mathbb{R}^{5|8}$ and $\mathbb{R}^{4|8}$ are:  
\begin{align}
    &\theta^{\hat{\alpha}}_1=(\theta^{\alpha}_1\;,\;\bar{\theta}_{2\dot{\alpha}})\;\;,\;\;\theta^{\hat{\alpha}}_2=(\theta^{\alpha}_2\;,\;-\bar{\theta}_{1\dot{\alpha}})\nn \\
    &\theta_{1\hat{\alpha}}=\binom{\theta_{1\alpha}}{\bar{\theta}^{\dot{\alpha}}_2}\;\;,\;\;\theta_{2\hat{\alpha}}=\binom{\theta_{2\alpha}}{-\bar{\theta}^{\dot{\alpha}}_1}
\end{align}
The corresponding covariant derivatives will be:
\begin{align}
    &D^{\hat{\alpha}}_1=(D^{\alpha}\;,\;\bar{Q}_{\dot{\alpha}})\;\;,\;\;D^{\hat{\alpha}}_2=(Q^{\alpha}\;,\;-\bar{D}_{1\dot{\alpha}})\nn \\
    &D_{1\hat{\alpha}}=\binom{D_{\alpha}}{\bar{Q}^{\dot{\alpha}}}\;\;,\;\;D_{2\hat{\alpha}}=\binom{Q_{\alpha}}{-\bar{D}^{\dot{\alpha}}}
\end{align}
Then, as we defined in (\ref{proj.theta}) the 5D projective Grassmann coordinates are in terms of 4D variables:
\begin{align}
    &\Theta^{\hat{\alpha}}=(\theta^\alpha_2-\zeta\theta^\alpha_1\;,\;-\bar{\theta}_{\dot{\alpha}1}-\zeta\bar{\theta}_{\dot{\alpha}2})\;\;,\;\;\Theta_{\hat{\alpha}}=\binom{\theta_{2\alpha}-\zeta\theta_{1\alpha}}{-\bar{\theta}^{\dot{\alpha}}_1-\zeta\bar{\theta}^{\dot{\alpha}}_2} \nn \\
    &\Theta^2_5=\frac{1}{2}\Theta^{\hat{\alpha}}\Theta_{\hat{\alpha}}=\theta^2_4+\bar{\theta}^2_4
\end{align}
Above we also defined the 4-dimensional projective Grassmann variables.\\
In this description the $\nabla$ and $\Delta$ operators are expressed with 4D pairs as:
\begin{align}
    &\nabla^{\hat{\alpha}}= D^{\hat{\alpha}}_1+\zeta D^{\hat{\alpha}}_2=(\nabla^\alpha_4\;,\,\bar{\nabla}_{4\dot{\alpha}})\;\;,\;\;\nabla_{\hat{\alpha}}= D_{1\hat{\alpha}}+\zeta D_{2\hat{\alpha}}=\binom{\nabla^\alpha_4}{\bar{\nabla}_{4\dot{\alpha}}} \label{5Dnablas}\\
    &\Delta^{\hat{\alpha}}=D^{\hat{\alpha}}_2-\frac{1}{\zeta}D^{\hat{\alpha}}_1=(\Delta^\alpha_4\;,\;-\bar{\Delta}_{4\dot{\alpha}})\;\;,\;\;\Delta_{\hat{\alpha}}=D_{\hat{\alpha}2}-\frac{1}{\zeta}D_{\hat{\alpha}1}=\binom{\Delta_{4\alpha}}{-\bar{\Delta}^{\dot{\alpha}}_4}
\end{align}
Based on these relations, we now obtain the 4D chiral/antichiral SYM from the action in (\ref{YM5D}). First, the field strength needs to be expressed by $\mathcal{W}_4$ and $\bar{\mathcal{W}}_4$. To do this, let us write the 5D superalgebra in terms of 2$\times$2 matrices.
\begin{align}
    \{\mathcal{D}^i_{\hat{\alpha}},\mathcal{D}^j_{\hat{\beta}}\}=i\epsilon^{ij}\left[
    \begin{pmatrix}
    i\epsilon_{\alpha\beta}\mathcal{D}_5 & \nabla_{\alpha\dot{\beta}}\\
    \nabla_{\dot{\alpha}\beta} & -i\epsilon_{\dot{\alpha}\dot{\beta}}\mathcal{D}_5
    \end{pmatrix}
    +\begin{pmatrix}
    \epsilon_{\alpha\beta} & 0 \\
    0 & \epsilon_{\dot{\alpha}\dot{\beta}}
    \end{pmatrix}\mathbb{W}_5
    \right]
\end{align}
After descending from 4-spinor to 2-spinor representations, the resulting anticommutators are equivalent to the 4D superalgebra:  
\begin{align}    
    &\{\mathcal{D}^i_{\alpha},\mathcal{D}^j_{\beta}\}=i\epsilon^{ij}(\nabla_5+\epsilon_{\alpha\beta}\mathbb{W}_5)=i\epsilon^{ij}\epsilon_{\alpha\beta}\bar{\mathbb{W}_4}\;;\\
    &\{\bar{\mathcal{D}}^i_{\dot{\alpha}},\bar{\mathcal{D}}^j_{\dot{\beta}}\}=i\epsilon^{ij}(-\nabla_5+\epsilon_{\dot{\alpha}\dot{\beta}}\mathbb{W}_5)=i\epsilon^{ij}\epsilon_{\dot{\alpha}\dot{\beta}}\mathbb{W}_4
\end{align}
Here we have expressed the gauge covariantized $\nabla_{\hat{\alpha}\hat{\beta}}$ along the 5-th spacetime direction using the 5D antisymmetric $\Gamma$ matrix:
\begin{align}
    (\nabla_5)_{\hat{\alpha}\hat{\beta}}=(\partial_5+A_5)_{\hat{\alpha}\hat{\beta}}=(C\Gamma^5)_{\hat{\alpha}\hat{\beta}}\mathcal{D}_5=\begin{pmatrix}
    i\epsilon_{\alpha\beta} & 0\\
    0 & -i\epsilon_{\dot{\alpha}\dot{\beta}}
    \end{pmatrix}\mathcal{D}_5
\end{align}
Here we used the chiral representation for $SO(4,1)$, where the gamma matrix is block diagonal and antisymmetric, acting differently on the left- and right-handed Weyl spinors.\\
\begin{equation}
    \{\mathcal{D}^i_{\hat{\alpha}},\mathcal{D}^j_{\hat{\beta}}\}=i\epsilon^{ij}
    \begin{pmatrix}
    i\epsilon_{\alpha\beta}(\mathcal{D}_5-i\mathbb{W}_5) & \nabla_{\alpha\dot{\beta}}\\
    \nabla_{\dot{\alpha}\beta}  & -i\epsilon_{\dot{\alpha}\dot{\beta}}(\mathcal{D}_5+i\mathbb{W}_5)
    \end{pmatrix}
    =\begin{pmatrix}
    \{\mathcal{D}^i_{\alpha},\mathcal{D}^j_{\beta}\} & \{\mathcal{D}^i_{\alpha},\bar{\mathcal{D}}^j_{\dot{\beta}}\}\\
    \{\bar{\mathcal{D}}^i_{\dot{\alpha}},\mathcal{D}^j_{\beta}\}  & \{\bar{\mathcal{D}}^i_{\dot{\alpha}},\bar{\mathcal{D}}^j_{\dot{\beta}}\}
    \end{pmatrix}
\end{equation}
Comparing the 5D and 4D algebras, we find:
\begin{align}
    i&\mathcal{D}_5+\mathbb{W}_5=\bar{\mathbb{W}}_4 \nn \\
    -i&\mathcal{D}_5+\mathbb{W}_5=\mathbb{W}_4
\end{align}
Then we have the 5D field strength in terms of 4D ones.
\begin{equation}
    \mathbb{W}_5=\frac{1}{2}(\mathbb{W}_4+\bar{\mathbb{W}}_4)
\end{equation}
It is worth to mention that 5D anticommuting covariant derivatives produce torsion term $\boldsymbol{\nabla}_{\hat{\alpha}\hat{\beta}}$ plus the field strength term $\mathbb{W}_5$ while in 4D, the anticommutation of only chiral or only antichiral covariant derivatives contain just the field strengths $\mathbb{W}_4$ and $\bar{\mathbb{W}}_4$ respectively, hence comparing the diagonal part of the 5D anticommutation matrix with the 4D anticommutation, we receive relations between the 5-th spacetime dimension covariant derivate, 5D field strenth and 4D field strengths.\\

Now that we know 5D field strength is expressed with 4D chiral field strength and its complex conjugation, we reduce the Yang-Mills action in (\ref{YM5D}). Projective, descendent field $G$ in 4D:
\begin{equation}
    G\vert_{4D}=\nabla_4^2\mathcal{W}_4^2+\bar{\nabla}_4^2\bar{\mathcal{W}}_4^2
\end{equation}
The 4-th degree of $\Delta_5$ reduces to:
\begin{equation}
    \Delta_5^4=\left(\Delta_4^2+\bar{\Delta}_4^2\right)^2=2\Delta_4^2\bar{\Delta}_4^2
\end{equation}
There will be spacetime derivatives when swapping $\Delta$ with $\bar{\Delta}$, however, we can ignore them, since they are integrated over the spacetime. 
\begin{align}
    S_{YM}=&\int d^4x \Delta^2\bar{\Delta}^2\left[(\theta^2+\bar{\theta}^2)(\nabla^2\mathcal{W}^2+\bar{\nabla}^2\bar{\mathcal{W}}^2)\right] \nn \\
    \sim &\int d^4x \frac{\partial^2}{\partial\theta^2}\frac{\partial^2}{\partial\bar{\theta}^2}\left[(\theta^2+\bar{\theta}^2)(\nabla^2\mathcal{W}^2+\bar{\nabla}^2\bar{\mathcal{W}}^2)\right]
\end{align}
Taking into account 4-dimensional chirality and the Bianchi identities, it is easy to see the above action reduces to just the sum of chiral/antichiral actions:
\begin{equation}
    S_{YM}\sim \int d^4x \Delta^2 \nabla^2\mathcal{W}^2 + \int d^4x \bar{\Delta}^2 \bar{\nabla}^2\bar{\mathcal{W}}^2
\end{equation}
This shows that the proposed projective formula for the 5-dimensional YM action (\ref{YM5D}) is correct.
\subsection{Discussion}
In this section, we see that 5-dimensional Yang-Mills action can be described in terms of field strength in projective superspace. We coupled a constant projective field depending on Grassmann coordinates with the projective covariant field which is the descendent derivative of the field strength. Because we defined the derivatives with respect to the projective Grassman coordinates as proportional to the $\Delta$ derivatives, it was straightforward to prove that this action reduces to the 4-dimensional chiral action and its complex conjugate part.  

%% file: sections/section2.tex
\section{Chern-Simons action}\label{sec:Chern-Simons}

We take a hint from the description of the variation of the 5-dimensional Chern-Simons action proposed by Zupnik in \cite{Zupnik:1999iy} and later corrected in the erratum \cite{ZUPNIK2002405}. The integrability of Zupnik's variational form is proven by Kuzenko in \cite{Kuzenko:2006ek}. In projective superspace, the action should contain $e^{-V}\delta e^V$ as the variation of the prepotential. This is because of the way vector multiplets couple to polar/hypermultiplets. By polar multiplets we mean the arctic: analytic around the north pole of the manifold $\mathbb{CP}^1$: $\Upsilon=\sum_{n=0}^\infty \Upsilon_n \zeta^n$ and the antarctic: analytic around the south pole $\bar{\Upsilon}=\sum_{n=0}^{\infty}\bar{\Upsilon}_{-n}\zeta^{-n}$. This form is also consistent with the overall gauge invariance of the action, as we will see below.
\begin{align} \label{G}
    \delta S_{CS}=k_5\tr \int d^5 x d^8\theta\oint d\zeta e^{-V}\delta e^V \{A\;,\;\mathcal{W}\}
\end{align}
The infinitesimal gauge transformation rule for the non-Abelian vector multiplet has the standard form. Remind you that the gauge parameters $i\bar{\Lambda}$, $i\Lambda$ are projective polar multiplets.
\begin{align}
    \delta e^V=i\bar{\Lambda}e^V-e^V i\Lambda\;\;,\;\;
    \delta A=-i\partial_\zeta \Lambda + [i\Lambda,A]\;\;,\;\; 
    \delta W=[i\Lambda,\mathcal{W}]
\end{align}
Additionally, useful to remember (\ref{useful}): $e^V A e^{-V}=\tilde{A}+\partial_\zeta e^V e^{-V}$ and $e^V\mathcal{W}e^{-V}=\widetilde{\mathcal{W}}$. None of the $i\bar{\Lambda}\tilde{A}\tilde{\mathcal{W}}$ and $i\Lambda A \mathcal{W}$ contains $\zeta^{-1}$ term, therefore, vanishes after contour integrals. Furthermore, one could easily check that $\nabla^4(i\bar{\Lambda}\tilde{\mathcal{W}}\partial_\zeta e^V e^{-V})=0$. Therefore,
\begin{align}
    \delta S_{\Lambda}=k_5\tr \int d^5 x d^8\theta\oint d\zeta \left( i\bar{\Lambda}\{\tilde{A}\;,\;\tilde{\mathcal{W}}\}+i\bar{\Lambda}\{\partial_\zeta e^V e^{-V}\;,\;\tilde{\mathcal{W}}\} -i\Lambda \{A\;,\;\mathcal{W}\}\right)=0
\end{align}
proven to be gauge-invariant.
\subsection{Abelian Chern-Simons action}
Now that we know that this formula (\ref{G}) for variation is gauge invariant, let us first derive the Abelian action from its corresponding variation. 
\begin{equation} \label{AbC-S}
    \delta S_{CS}^{Ab}=k_5\tr \int d^5 x d^8\theta\oint d\zeta \delta V A\nabla^2 A
\end{equation}
where the Abelian gauge connection $A_\zeta$ is as we established in the last section of the article \cite{Davgadorj:2017ezp}:
\begin{equation}
    A_{\zeta_0}=\oint d\zeta_1 \frac{V_1}{\zeta_{10}^2}
\end{equation}
where $V(\zeta)$ is Abelian vector prepotential, projective multiplet: $V(\zeta)=\sum_{n=-\infty}^\infty v_n \zeta^n$. We will discuss the physical degrees of freedom contained in this prepotential multiplet in the end when we check the consistency of the obtained non-Abelian action reduced to the Abelian case.\\
Then, a partial derivative can be pulled from the variation on (\ref{AbC-S}).
\begin{align}
    (\ref{AbC-S})=&\delta \;k_5 \int d^5 x d^8\theta\oint d\zeta_0 V_0 \oint d\zeta_1 \frac{V_1}{\zeta_{10}^2}\nabla_0^2 \oint d\zeta_2 \frac{V_2}{\zeta_{20}^2}+ \\
    &-2 k_5 \int d^5 x d^8\theta\oint d\zeta_0 V_0 \oint d\zeta_1 \frac{\delta V_1}{\zeta_{10}^2}\nabla_0^2 \oint d\zeta_2 \frac{V_2}{\zeta_{20}^2} \label{Ab_sec.t}
\end{align}
Here we can apply renaming $\zeta_0\longleftrightarrow\zeta_1$ to the second line and there reorder the first and the second parts 
\begin{align}
    -2 k_5 \int d^5 x d^8\theta\oint d\zeta_0 \delta V_0 \oint d\zeta_1 \frac{V_1}{\zeta_{01}^2}\nabla_1^2 \oint d\zeta_2 \frac{V_2}{\zeta_{21}^2}
\end{align}
Again, recalling the 4-dimensional chapter, we use the following relation:
\begin{equation}
    \oint d\zeta_1 \frac{V_1}{\zeta_{01}^2}=-\partial_{\zeta_0}V_0+\oint d\zeta_1 \frac{V_1}{\zeta_{10}^2}
\end{equation}
Another known fact from the 4-dimensions for any projective function $\mathcal{P}$:
\begin{equation}
      \nabla^2 \partial_{\zeta} \mathcal{P}=0
\end{equation}
The same is here in 5-dimensions, leading to the conclusion, saying that $\nabla^2$ acting on positive or negative projections of the $\zeta$-derivative of any projective field is $\zeta$-independent.
\begin{equation} \label{Ab_W}
    \mathcal{W}^{Ab}=\nabla_1^2\oint d\zeta_2 \frac{V_2}{\zeta_{21}^2}=\nabla_1^2\oint d\zeta_2 \frac{V_2}{\zeta_{12}^2}=D_2^{\hat{\alpha}}D_{2\hat{\alpha}}v_{-1}=D_1^{\hat{\alpha}}D_{1\hat{\alpha}}v_{1}
\end{equation}
Therefore, the second line (\ref{Ab_sec.t}) is reduced to the same form as the original variation itself (\ref{AbC-S}). This allows us to find the Abelian action:
\begin{align} \label{5dAbCS}
    3\delta S_{CS}^{Ab}=&3 k_5\int d^5 x d^8\theta \oint d\zeta_0 \delta V_0 \oint d\zeta_1 \frac{V_1}{\zeta_{10}^2}\nabla_0^2 \oint d\zeta_2 \frac{V_2}{\zeta_{20}^2} \nn \\
    &=\delta k_5\int d^5 x d^8\theta \oint d\zeta_0 V_0 \oint d\zeta_1 \frac{V_1}{\zeta_{10}^2}\nabla_0^2 \oint d\zeta_2 \frac{V_2}{\zeta_{20}^2} \nn \\
    S_{CS}^{Ab}=&\frac{1}{3}  k_5\int d^5 x d^8\theta \oint d\zeta_0 V_0 A_{\zeta_0}\nabla_0^2 A_{\zeta_0}
\end{align} 

This form of the Abelian action is in agreement with the Zupnik's action analogously.\\ 

In this derivation, the key simplification comes from the $\zeta$-independent description of the Abelian field strength (\ref{Ab_W}). Can we also describe the non-Abelian field strength in a similar manner as a $\zeta$-dependent part $\times$ $\zeta$-indepent term $\times$ $\zeta$-dependent part?\\ 

In fact, we used a similar description in 4-dimensions to prove the equivalence of the chiral action and the full-measure action \cite{Davgadorj:2023lfv}. We split the prepotential into nonsymmetric parts. In 5-dimensions this splitting will lead us to find desirable expressions for the field strength as we needed.
\begin{align} \label{relation:W:F}
    \mathcal{W}=&e^{-\check{U}}(D_1^{\hat{\alpha}}D_{1\hat{\alpha}}\bar{F}) e^{\check{U}}\\
    \mathcal{W}=&e^{-\hat{U}}(D_2^{\hat{\alpha}}D_{2\hat{\alpha}} F) e^{\hat{U}}
\end{align}
Here, $F$ (\ref{F}) and $\bar{F}$ (\ref{barF}) are Hermitian conjugate to each other and additionally related through $e^Pe^{\bar{P}}$ (\ref{eP}) function: 
\begin{equation}
    D_1^{\hat{\alpha}}D_{1\hat{\alpha}}\bar{F}=e^{-\bar{P}}e^{-P}(D_2^{\hat{\alpha}}D_{2\hat{\alpha}} F)e^Pe^{\bar{P}}
\end{equation}
For the definitions of these functions and techniques on how to obtain them one can refer to the appendices.

\subsection{Non-Abelian Chern-Simons action}
Our task now is to derive the actual non-Abelian action from the proposed variation form (\ref{G}). Let us first notice that for this variation we can equivalently write as 
\begin{align} \label{K} 
    \delta S_{CS}=-k_5\tr \int d^5 x d^8\theta\oint d\zeta e^{-V}\delta e^V \nabla^{\hat{\alpha}}A\nabla_{\hat{\alpha}}A
\end{align}
We can define here the projective fields $G$ and $K$. In fact, the construction (\ref{G}) stems from first constructing the projective covariant supefield $G$ and then coupling it to the vector $e^{-V}\delta e^V$; finally, the whole coupling comes with the projective measure $\Delta^4$.  

 \begin{align} 
   & G=\nabla^4\{A\;,\;\mathcal{W}\}=2\left(\nabla^2\mathcal{W}^2+\nabla^{\hat{\alpha}}\mathcal{W}\nabla_{\hat{\alpha}}\mathcal{W}\right) \\
   & K=\nabla^4\left(\nabla^{\hat{\alpha}}A\nabla_{\hat{\alpha}}A\right)=-2\nabla^2\mathcal{W}^2+3\nabla^{\hat{\alpha}}\mathcal{W}\nabla_{\hat{\alpha}}\mathcal{W}
 \end{align}

The full measure of the action is $\mathrm{d}^8\theta\thicksim\Delta^4\nabla^4$, then since the Lagrangian $e^{-V}\delta e^V G$ is a projective, we can use the projective half measure: $\frac{\left(D^{\hat{\alpha}}_2D_{1\hat{\alpha}}\right)^2}{\zeta^2}$ instead of $\Delta^4$. This comes from the fact that $\Delta$'s can be written in terms of $\nabla$'s and the $D$'s as:
\begin{equation}
    \Delta^{\hat{\alpha}}=2D^{\hat{\alpha}}_2-\frac{1}{\zeta}\nabla^{\hat{\alpha}}\;\;,\;\;\Delta_{\hat{\alpha}}=-\frac{2}{\zeta}D_{1\hat{\alpha}}+\frac{1}{\zeta}\nabla_{\hat{\alpha}}
\end{equation}
Then $\Delta^4\nabla^4=(\frac{1}{2}\Delta^{\hat{\alpha}}\Delta_{\hat{\alpha}})^2\nabla^4=\frac{4}{\zeta^2}(D^{\hat{\alpha}}_2D_{1\hat{\alpha}})^2\nabla^4$. Hence, the variation is:
\begin{align}
    \delta S_{CS}=4k_5\tr \int d^5 x \left(D^{\hat{\alpha}}_2D_{1\hat{\alpha}}\right)^2\oint \frac{d\zeta}{\zeta^2}\left(e^{-V}\delta e^V G\right)
\end{align}
Since (\ref{G}) and (\ref{K}) (notice the negative sign) are equivalent, one can imply:
\begin{align}\label{contra=zero}
     \tr \int d^5 x \oint d\zeta \Delta^4 \left(e^{-V}\delta e^V \nabla^{\hat{\alpha}}\mathcal{W}\nabla_{\hat{\alpha}}\mathcal{W}\right)=0
\end{align}
This means consequently we can write the variation further:
\begin{align}
    \delta S_{CS}=8k_5\tr \int d^5 x \left(D^{\hat{\alpha}}_2D_{1\hat{\alpha}}\right)^2 \oint \frac{\nabla^2}{\zeta^2} \left(e^{-V}\delta e^V \mathcal{W}^2\right)
\end{align} 
Let us now rewrite the integrand using the decomposition $e^V=e^{\hat{\bar{U}}}e^{\check{U}}$ and remind you that in this representation (we will introduce this decomposition in the appendices) the field strength is expressed as $\mathcal{W}=e^{-\check{U}}\left(D^{\hat{\alpha}}_1D_{1\hat{\alpha}}\bar{F}\right)e^{\check{U}}$. Then we can write the following:
\noindent
\begin{align}
    &\oint d\zeta \frac{\left(D^{\hat{\alpha}}_2D_{1\hat{\alpha}}\right)^2}{\zeta^2}\nabla^2 \left[e^{-\hat{\bar{U}}}\delta e^{\hat{\bar{U}}}\left(D^{\hat{\alpha}}_1D_{1\hat{\alpha}}\bar{F}\right)^2+\delta e^{\check{U}}e^{-\check{U}}\left(D^{\hat{\alpha}}_1D_{1\hat{\alpha}}\bar{F}\right)^2\right]=\\ 
    =&\oint d\zeta\frac{\left(D^{\hat{\alpha}}_2D_{1\hat{\alpha}}\right)^2}{\zeta^2}\left\{ \zeta\left(D^{\hat{\alpha}}_2D_{1\hat{\alpha}}\right) \left[e^{-\bar{P}}e^{-P}\delta(e^P e^{\bar{P}})\left(D^{\hat{\alpha}}_1D_{1\hat{\alpha}}\bar{F}\right)^2\right]+\right. \nonumber \\ 
    +&\frac{1}{2}\zeta^2\left(D^{\hat{\alpha}}_2D_{2\hat{\alpha}}\right)\left[\frac{N}{\zeta}\delta(e^Pe^{\bar{P}})\left(D^{\hat{\alpha}}_1D_{1\hat{\alpha}}\bar{F}\right)^2+(e^{-\bar{P}}e^{-P})\frac{\delta M}{\zeta}\left(D^{\hat{\alpha}}_1D_{1\hat{\alpha}}\bar{F}\right)^2\right] \nonumber\\ 
    +&\left.\frac{1}{2}\left(D^{\hat{\alpha}}_1D_{1\hat{\alpha}}\right) \left[\delta \bar{F}\zeta \left(D^{\hat{\alpha}}_1D_{1\hat{\alpha}}\bar{F}\right)^2\right]\right\}
\end{align}
Here, we have rewritten $\nabla^2$ in terms of $\zeta$'s:
\begin{align}
    \nabla^2=\frac{1}{2}D^{\hat{\alpha}}_1D_{1\hat{\alpha}}+\zeta D^{\hat{\alpha}}_2D_{1\hat{\alpha}}+\frac{1}{2}\zeta^2 D^{\hat{\alpha}}_2D_{2\hat{\alpha}}
\end{align}
$N$ and $M$ are so far not defined $\zeta$-independent functions. But we do not need to worry about them, since we will replace this term with the convenient $F$ function later. However, for the sake of consistency, let us write them down. $F$, $\bar{F}$ and $e^Pe^{\bar{P}}$ and its inverse are defined respectively in (\ref{F}), (\ref{barF}) and (\ref{eP}), (\ref{inveP}).
\begin{align}
    &N=\sum^\infty_{n=1}(-1)^{n}\oint d\zeta_1\dots d\zeta_n \frac{X_1\dots X_n}{\zeta_{21}\dots \zeta_{n,n-1}}\frac{\zeta_n}{\zeta_1} \\
    &M=\sum^\infty_{n=1}(-1)^{n+1}\oint d\zeta_1\dots d\zeta_n \frac{X_1\dots X_n}{\zeta_{21}\dots \zeta_{n,n-1}}\frac{\zeta_1}{\zeta_n}
\end{align}
Equivalently, we could also apply the decomposition $e^V=e^{\check{\bar{U}}}e^{\hat{U}}$, and in this splitting, the field strength is written as $\mathcal{W}=e^{-\hat{U}}\left(D^{\hat{\alpha}}_2D_{2\hat{\alpha}}F\right)e^{\hat{U}}$. Then:
\begin{align}
    &\oint d\zeta \frac{\left(D^{\hat{\alpha}}_2D_{1\hat{\alpha}}\right)^2}{\zeta^2}\nabla^2 \left[e^{-\check{\bar{U}}}\delta e^{\check{\bar{U}}}\left(D^{\hat{\alpha}}_2D_{2\hat{\alpha}}F\right)^2+\delta e^{\hat{U}}e^{-\hat{U}}\left(D^{\hat{\alpha}}_2D_{2\hat{\alpha}}F\right)^2\right]=\\ 
    =&\oint d\zeta\frac{\left(D^{\hat{\alpha}}_2D_{1\hat{\alpha}}\right)^2}{\zeta^2}\left\{ \zeta\left(D^{\hat{\alpha}}_2D_{1\hat{\alpha}}\right) \left[\delta(e^P e^{\bar{P}})e^{-\bar{P}}e^{-P}\left(D^{\hat{\alpha}}_2D_{2\hat{\alpha}}F\right)^2\right]+\right. \nonumber \\ 
    +&\frac{1}{2}\left(D^{\hat{\alpha}}_1D_{1\hat{\alpha}}\right)\left[\zeta\delta S(e^{-\bar{P}}e^{-P})\left(D^{\hat{\alpha}}_2D_{2\hat{\alpha}}F\right)^2+\zeta\delta(e^Pe^{\bar{P}})T\left(D^{\hat{\alpha}}_2D_{2\hat{\alpha}}F\right)^2\right] \nonumber\\ 
    +&\left.\frac{1}{2}\zeta^2\left(D^{\hat{\alpha}}_2D_{2\hat{\alpha}}\right) \left[\frac{\delta F}{\zeta} \left(D^{\hat{\alpha}}_2D_{2\hat{\alpha}}F\right)^2\right]\right\}
\end{align}
$S$ and $T$ are not important here, but we give them away anyways.
\begin{align}
    &S=\sum^\infty_{n=1}(-1)^{n+1}\oint d\zeta_1\dots d\zeta_n \frac{X_1\dots X_n}{\zeta_{21}\dots \zeta_{n,n-1}}\frac{1}{\zeta_n^2} \\
    &T=\sum^\infty_{n=1}(-1)^{n}\oint d\zeta_1\dots d\zeta_n \frac{X_1\dots X_n}{\zeta_{21}\dots \zeta_{n,n-1}}\frac{1}{\zeta_1^2}
\end{align}
Now, those two expressions (using different decompositions) for the variation of the action are obviously equal, and therefore we can compare their terms with the same measure. Using the relation
\begin{equation}
\left(D^{\hat{\alpha}}_2D_{2\hat{\alpha}}F\right)=e^P e^{\bar{P}}\left(D^{\hat{\alpha}}_1D_{1\hat{\alpha}}\bar{F}\right)e^{-\bar{P}}e^{-P}
\end{equation}
we can see that both the first lines with the $\left(D^{\hat{\alpha}}_2D_{1\hat{\alpha}}\right)$ derivatives are indeed the same expression.
Comparing the other two lines, we realise the following: 
\begin{align}
    &\int d^5x\left(D^{\hat{\alpha}}_1D_{1\hat{\alpha}}\right)\left[\delta S(e^{-\bar{P}}e^{-P})\left(D^{\hat{\alpha}}_2D_{2\hat{\alpha}}F\right)^2+\delta(e^Pe^{\bar{P}})T\left(D^{\hat{\alpha}}_2D_{2\hat{\alpha}}F\right)^2\right]= \nonumber \\
    =&\int d^5x\left(D^{\hat{\alpha}}_1D_{1\hat{\alpha}}\right) \left[\delta \bar{F} \left(D^{\hat{\alpha}}_1D_{1\hat{\alpha}}\bar{F}\right)^2\right] \\
    &\int d^5x\left(D^{\hat{\alpha}}_2D_{2\hat{\alpha}}\right)\left[N \delta (e^Pe^{\bar{P}})\left(D^{\hat{\alpha}}_1D_{1\hat{\alpha}}\bar{F}\right)^2+(e^{-\bar{P}}e^{-P})\delta M\left(D^{\hat{\alpha}}_1D_{1\hat{\alpha}}\bar{F}\right)^2\right]= \nonumber \\
    =&\int d^5x\left(D^{\hat{\alpha}}_2D_{2\hat{\alpha}}\right) \left[\delta F \left(D^{\hat{\alpha}}_2D_{2\hat{\alpha}}F\right)^2\right]
\end{align}
Therefore, we can use only the $F$ and $\bar{F}$ expressions as:
\begin{align}\label{CSvariation_pr.measure}
    &\delta S_{CS}=8k_5\tr \int d^5 \left(D^{\hat{\alpha}}_2D_{1\hat{\alpha}}\right)^2\left\{ \left(D^{\hat{\alpha}}_2D_{1\hat{\alpha}}\right) \left[\delta(e^P e^{\bar{P}})e^{-\bar{P}}e^{-P}\left(D^{\hat{\alpha}}_2D_{2\hat{\alpha}}F\right)^2\right]+\right. \nonumber \\ 
    +&\left.\frac{1}{2}\left(D^{\hat{\alpha}}_1D_{1\hat{\alpha}}\right)\left[\delta \bar{F}\left(D^{\hat{\alpha}}_1D_{1\hat{\alpha}}\bar{F}\right)^2\right]+\frac{1}{2}\left(D^{\hat{\alpha}}_2D_{2\hat{\alpha}}\right) \left[\delta F \left(D^{\hat{\alpha}}_2D_{2\hat{\alpha}}F\right)^2\right]\right\}
\end{align}

One could have received the above three terms with different numerical factors if they had started with the more general measure $\Delta^4$, instead of switching to the projective measure $\frac{(D^{\hat{\alpha}}_2D_{1\hat{\alpha}})^2}{\zeta^2}$. Remembering that the projective function $G$ is reduced to the form $\nabla^2\mathcal{W}^2$, we write the Chern-Simons variation in a general form:
\begin{align}
    \delta S_{CS}=k_5\tr \int d^5 x  \oint \Delta^4\nabla^2 \left(e^{-V}\delta e^V \mathcal{W}^2\right)
\end{align} 
Let us now find the component covariant derivative form of the measure $\Delta^4\nabla^2$. Before that we establish that with the space-time integration the following is true:
\begin{align}
    &\int d^5x \left(D^{\hat{\alpha}}_2D_{1\hat{\alpha}}\right) \left(D^{\hat{\alpha}}_2D_{2\hat{\alpha}}\right)=\int d^5x \left(D^{\hat{\alpha}}_2D_{2\hat{\alpha}}\right) \left(D^{\hat{\alpha}}_2D_{1\hat{\alpha}}\right) \nonumber \\
    &\int d^5x \left(D^{\hat{\alpha}}_1D_{1\hat{\alpha}}\right) \left(D^{\hat{\alpha}}_2D_{2\hat{\alpha}}\right)=\int d^5x \left(D^{\hat{\alpha}}_2D_{2\hat{\alpha}}\right) \left(D^{\hat{\alpha}}_1D_{1\hat{\alpha}}\right)
\end{align}
Taking those into account the measure becomes:
\begin{align}\label{long_measure}
    \Delta^4\nabla^2=&\frac{3}{8}\left(D^{\hat{\alpha}}_1D_{1\hat{\alpha}}\right)\left(D^{\hat{\alpha}}_2D_{2\hat{\alpha}}\right)^2-\frac{1}{2}\left(D^{\hat{\alpha}}_2D_{2\hat{\alpha}}\right)\left(D^{\hat{\alpha}}_2D_{1\hat{\alpha}}\right)^2+ \nonumber\\
    &+\frac{1}{\zeta}\left[\left(D^{\hat{\alpha}}_2D_{1\hat{\alpha}}\right)^3-\frac{1}{2}\left(D^{\hat{\alpha}}_2D_{1\hat{\alpha}}\right)\left(D^{\hat{\alpha}}_2D_{2\hat{\alpha}}\right)\left(D^{\hat{\alpha}}_1D_{1\hat{\alpha}}\right)\right]+\nonumber\\
    &+\frac{1}{\zeta^2}\left[\frac{3}{8}\left(D^{\hat{\alpha}}_2D_{2\hat{\alpha}}\right)\left(D^{\hat{\alpha}}_1D_{1\hat{\alpha}}\right)^2-\frac{1}{2}\left(D^{\hat{\alpha}}_1D_{1\hat{\alpha}}\right)\left(D^{\hat{\alpha}}_2D_{1\hat{\alpha}}\right)^2\right]
\end{align}
Further, we can investigate the relations between these terms using the fact that $\nabla^6=0$.
\begin{align}\label{relations}
    &0=\int d^5x\oint\frac{d\zeta}{\zeta^3}\nabla^6=\int d^5x\left[\left(D^{\hat{\alpha}}_2D_{1\hat{\alpha}}\right)^3+\frac{6}{4}\left(D^{\hat{\alpha}}_2D_{2\hat{\alpha}}\right)\left(D^{\hat{\alpha}}_1D_{1\hat{\alpha}}\right)\left(D^{\hat{\alpha}}_2D_{1\hat{\alpha}}\right)\right]\nonumber\\
    &0=\int d^5x \oint\frac{d\zeta}{\zeta^4}\nabla^6=\int d^5x\left[\frac{3}{8}\left(D^{\hat{\alpha}}_1D_{1\hat{\alpha}}\right)\left(D^{\hat{\alpha}}_2D_{2\hat{\alpha}}\right)^2+\frac{3}{2}\left(D^{\hat{\alpha}}_2D_{2\hat{\alpha}}\right)\left(D^{\hat{\alpha}}_2D_{1\hat{\alpha}}\right)^2\right] \nonumber\\
    &0=\int d^5x \oint\frac{d\zeta}{\zeta^2}\nabla^6=\int d^5x\left[\frac{3}{8}\left(D^{\hat{\alpha}}_2D_{2\hat{\alpha}}\right)\left(D^{\hat{\alpha}}_1D_{1\hat{\alpha}}\right)^2+\frac{3}{2}\left(D^{\hat{\alpha}}_1D_{1\hat{\alpha}}\right)\left(D^{\hat{\alpha}}_2D_{1\hat{\alpha}}\right)^2\right]
\end{align}
Once we write the integrand $e^{-V}\delta e^V \mathcal{W}^2$ using non-symmetric splitting and the whole expression becomes $\zeta$-independent, just as we did earlier with the projective measure, we can then use the relations (\ref{relations}) to simplify the measure in (\ref{long_measure}). This will give us the Chern-Simons variation:
\begin{align}\label{CSvariation_long.measure}
    &\delta S_{CS}=2k_5\tr \int d^5\left\{ \frac{4}{3}\left(D^{\hat{\alpha}}_2D_{1\hat{\alpha}}\right)^3\left[\delta(e^P e^{\bar{P}})e^{-\bar{P}}e^{-P}\left(D^{\hat{\alpha}}_2D_{2\hat{\alpha}}F\right)^2\right]\right. \nonumber \\ 
    &\left.-2\left(D^{\hat{\alpha}}_2D_{1\hat{\alpha}}\right)^2\left(D^{\hat{\alpha}}_1D_{1\hat{\alpha}}\right)\left[\delta \bar{F}\left(D^{\hat{\alpha}}_1D_{1\hat{\alpha}}\bar{F}\right)^2\right]-2\left(D^{\hat{\alpha}}_2D_{1\hat{\alpha}}\right)^2\left(D^{\hat{\alpha}}_2D_{2\hat{\alpha}}\right) \left[\delta F \left(D^{\hat{\alpha}}_2D_{2\hat{\alpha}}F\right)^2\right]\right\}
\end{align}
Two forms of variation (\ref{CSvariation_pr.measure}) and (\ref{CSvariation_long.measure}) are equal. Therefore, we can find the term with the $\left(D^{\hat{\alpha}}_2D_{1\hat{\alpha}}\right)^3$ measure from this equality.
\begin{align}
     &\tr \int d^5 x\left(D^{\hat{\alpha}}_2D_{1\hat{\alpha}}\right)^3\left[\delta(e^P e^{\bar{P}})e^{-\bar{P}}e^{-P}\left(D^{\hat{\alpha}}_2D_{2\hat{\alpha}}F\right)^2\right]= \nonumber\\
     &=-\frac{3}{2}\tr \int d^5x \left(D^{\hat{\alpha}}_2D_{1\hat{\alpha}}\right)^2\left\{\left(D^{\hat{\alpha}}_1D_{1\hat{\alpha}}\right)\left[\delta \bar{F}\left(D^{\hat{\alpha}}_1D_{1\hat{\alpha}}\bar{F}\right)^2\right]+\left(D^{\hat{\alpha}}_2D_{2\hat{\alpha}}\right) \left[\delta F \left(D^{\hat{\alpha}}_2D_{2\hat{\alpha}}F\right)^2\right]\right\}
\end{align}
Then, we replace it in the variation, so that we now have only two analogous terms.
\begin{align} \label{before the finale}
    &\delta S_{CS}=-8k_5 \tr \int d^5x \left(D^{\hat{\alpha}}_2D_{1\hat{\alpha}}\right)^2\left\{\left(D^{\hat{\alpha}}_1D_{1\hat{\alpha}}\right)\left[\delta \bar{F}\left(D^{\hat{\alpha}}_1D_{1\hat{\alpha}}\bar{F}\right)^2\right]+\left(D^{\hat{\alpha}}_2D_{2\hat{\alpha}}\right) \left[\delta F \left(D^{\hat{\alpha}}_2D_{2\hat{\alpha}}F\right)^2\right]\right\} \nonumber\\
    =&2k_5\tr \int d^5x \left(D^{\hat{\alpha}}_2D_{2\hat{\alpha}}\right)\left(D^{\hat{\alpha}}_1D_{1\hat{\alpha}}\right)\left\{\left(D^{\hat{\alpha}}_1D_{1\hat{\alpha}}\right)\left[\delta \bar{F}\left(D^{\hat{\alpha}}_1D_{1\hat{\alpha}}\bar{F}\right)^2\right]+\left(D^{\hat{\alpha}}_2D_{2\hat{\alpha}}\right) \left[\delta F \left(D^{\hat{\alpha}}_2D_{2\hat{\alpha}}F\right)^2\right]\right\}
\end{align}
Now, we treat these two terms separately, taking into account the fact that their measures already contain four $D_1$'s or four $D_2$'s for the second term. This fact leads to simple observations using partial integrations by $D_1$ (or $D_2$): 
\begin{align}
    \left(D_2D_2\right)\left(D_1D_1\right)^2\left[D_1D_1\bar{F}\cdot D^{\hat{\alpha}}_1\Bar{F}\cdot D_{1\hat{\alpha}}\delta \Bar{F}\right]=\left(D_2D_2\right)\left(D_1D_1\right)^2&\left[\frac{1}{4}D^{\hat{\alpha}}_1\Bar{F}\cdot D_1D_1\bar{F}\cdot D_{1\hat{\alpha}}\delta \Bar{F}+\right. \nonumber\\ 
    &\left.+\frac{1}{4}D^{\hat{\alpha}}_1\Bar{F}\cdot D_{1\hat{\alpha}}\Bar{F}\cdot D_1D_1\delta\Bar{F}\right] \\
    \left(D_2D_2\right)\left(D_1D_1\right)^2\left[D^{\hat{\alpha}}_1\Bar{F}\cdot D_1D_1\bar{F}\cdot D_{1\hat{\alpha}}\delta \Bar{F}\right]=\left(D_2D_2\right)\left(D_1D_1\right)^2&\left[\frac{1}{4}D_1D_1\bar{F}\cdot D^{\hat{\alpha}}_1\Bar{F}\cdot D_{1\hat{\alpha}}\delta \Bar{F}+\right. \nonumber\\ 
    &\left.+\frac{1}{4}D^{\hat{\alpha}}_1\Bar{F}\cdot D_{1\hat{\alpha}}\Bar{F}\cdot D_1D_1\delta\Bar{F}\right]
\end{align}
Those two relations are independent, therefore, substituting we find:
\begin{align} \label{intermediate}
   &\left(D_2D_2\right)\left(D_1D_1\right)^2\left[D_1D_1\bar{F}\cdot D^{\hat{\alpha}}_1\Bar{F}\cdot D_{1\hat{\alpha}}\delta \Bar{F}\right]=\frac{1}{3}\left(D_2D_2\right)\left(D_1D_1\right)^2\left[D^{\hat{\alpha}}_1\Bar{F}\cdot D_{1\hat{\alpha}}\Bar{F}\cdot D_1D_1\delta \Bar{F}\right] \nonumber\\
   &\left(D_2D_2\right)\left(D_1D_1\right)^2\left[D^{\hat{\alpha}}_1\Bar{F}\cdot D_1D_1\bar{F}\cdot D_{1\hat{\alpha}}\delta \Bar{F}\right]=\frac{1}{3}\left(D_2D_2\right)\left(D_1D_1\right)^2\left[D^{\hat{\alpha}}_1\Bar{F}\cdot D_{1\hat{\alpha}}\Bar{F}\cdot D_1D_1\delta \Bar{F}\right]
\end{align}
Further, we use the above 2 relations in the next relation repeating partial integrations:
\begin{align}
    &\left(D_2D_2\right)\left(D_1D_1\right)^2\left[D^{\hat{\alpha}}_1\Bar{F}\cdot D_{1\hat{\alpha}}\Bar{F}\cdot D_1D_1\delta\Bar{F}\right]=\left(D_2D_2\right)\left(D_1D_1\right)^2\left[D_1D_1\left(D^{\hat{\alpha}}_1\Bar{F}\cdot D_{1\hat{\alpha}}\Bar{F}\right)\cdot \delta\Bar{F}\right]= \nonumber\\
    &\left(D_2D_2\right)\left(D_1D_1\right)^2\left[D^{\hat{\alpha}}_1 D_1D_1\bar{F}\cdot D_{1\hat{\alpha}}\bar{F}\cdot \delta\Bar{F}+D^{\hat{\alpha}}_1\bar{F}\cdot D_{1\hat{\alpha}}D_1D_1\Bar{F}\cdot\delta \Bar{F}-2D^{\hat{\beta}}_1D^{\hat{\alpha}}_1\Bar{F}\cdot D_{1\hat{\beta}}D_{1\hat{\alpha}}\Bar{F}\cdot \delta\Bar{F}\right]=\nonumber\\
    &\left(D_2D_2\right)\left(D_1D_1\right)^2\left[-\frac{5}{2}(D_1D_1\Bar{F})^2\cdot \delta\Bar{F}-D_1D_1\Bar{F}\cdot D^{\hat{\alpha}}_1\Bar{F}\cdot D_{1\hat{\alpha}} \delta\Bar{F} -D^{\hat{\alpha}}_1\Bar{F}\cdot D_1D_1\Bar{F}\cdot D_{1\hat{\alpha}}\delta\Bar{F} \right]= \nonumber\\
    &\left(D_2D_2\right)\left(D_1D_1\right)^2\left[-\frac{5}{2}(D_1D_1\Bar{F})^2\cdot \delta\Bar{F}-\frac{2}{3}D^{\hat{\alpha}}_1\Bar{F}\cdot D_{1\hat{\alpha}}\Bar{F}\cdot D_1D_1\delta\Bar{F}\right]
\end{align}
This brings us to the next relation:
\begin{align}
    \left(D_2D_2\right)\left(D_1D_1\right)^2\left[(D_1D_1\Bar{F})^2\cdot \delta\Bar{F}\right]=-\frac{2}{3} \left(D_2D_2\right)\left(D_1D_1\right)^2\left[D^{\hat{\alpha}}_1\Bar{F}\cdot D_{1\hat{\alpha}}\Bar{F}\cdot D_1D_1\delta\Bar{F}\right]
\end{align}
On the other hand we can use the variation with respect to prepotentials $X_i$.
\begin{align}
    &\left(D_2D_2\right)\left(D_1D_1\right)^2\left[D^{\hat{\alpha}}_1\Bar{F}\cdot D_{1\hat{\alpha}}\Bar{F}\cdot D_1D_1\delta\Bar{F}\right]=\delta \left(D_2D_2\right)\left(D_1D_1\right)^2\left[D^{\hat{\alpha}}_1\Bar{F}\cdot D_{1\hat{\alpha}}\Bar{F}\cdot D_1D_1\Bar{F}\right]\nonumber\\
    &-\left(D_2D_2\right)\left(D_1D_1\right)^2\left[D_1D_1\bar{F}\cdot D^{\hat{\alpha}}_1\Bar{F}\cdot D_{1\hat{\alpha}}\delta \Bar{F}+D^{\hat{\alpha}}_1\Bar{F}\cdot D_1D_1\bar{F}\cdot D_{1\hat{\alpha}}\delta \Bar{F}\right] 
\end{align}
And then apply here the relations in (\ref{intermediate}) to find:
\begin{align}
    \left(D_2D_2\right)\left(D_1D_1\right)^2\left[D^{\hat{\alpha}}_1\Bar{F}\cdot D_{1\hat{\alpha}}\Bar{F}\cdot D_1D_1\delta\Bar{F}\right]=\frac{3}{5}\delta \left(D_2D_2\right)\left(D_1D_1\right)^2\left[D^{\hat{\alpha}}_1\Bar{F}\cdot D_{1\hat{\alpha}}\Bar{F}\cdot D_1D_1\Bar{F}\right]
\end{align}
This allows us to write:
\begin{align}
    \left(D_2D_2\right)\left(D_1D_1\right)^2\left[(D_1D_1\Bar{F})^2\cdot \delta \Bar{F}\right]&=-\frac{2}{5} \delta \left(D_2D_2\right)\left(D_1D_1\right)^2\left[D^{\hat{\alpha}}_1\Bar{F}\cdot D_{1\hat{\alpha}}\Bar{F}\cdot D_1D_1\Bar{F}\right]= \nonumber\\
    =&\frac{2}{5} \delta \left(D_2D_2\right)\left(D_1D_1\right)^2\left[\Bar{F}\cdot (D_1D_1\Bar{F})^2-\frac{1}{2}\Bar{F}^2\cdot (D_1D_1)^2\Bar{F}\right]
\end{align}

Since we can exchange ($D_1 \leftrightarrow D_2, F\leftrightarrow \Bar{F}$) and employ the same steps as above, the non-Abelian Chern-Simons action finally takes the form: 

\begin{align} \label{NonAb_CS}
   S_{CS}=\frac{1}{5}k_5 \tr \int d^5x \left(D_2D_2\right)\left(D_1D_1\right)&\left\{(D_1D_1)\left[\Bar{F}\cdot (D_1D_1\Bar{F})^2-\frac{1}{2}\Bar{F}^2\cdot (D_1D_1)^2\Bar{F}\right]+\right.\nonumber\\
   &\left.+(D_2D_2)\left[F\cdot (D_2D_2 F)^2-\frac{1}{2} F^2\cdot (D_2D_2)^2 F\right] \right\}
\end{align}

This expression can also be written in a manifestly $SU(2)$ covariant form in terms of projective covariant derivatives. It is easier to see the covariance of the $\nabla$ operators under $SU(2)$ transformation from their isospinor construction \ref{isospinor}.

\begin{align} \label{projective_NonAb}
   S_{CS}=\frac{8}{5}k_5 \tr \int d^5x \oint d\zeta \Delta^4 \nabla^2&\left\{\zeta\left[\Bar{F}\cdot (\nabla^2 \Bar{F})^2-\frac{1}{2}\Bar{F}^2\cdot \nabla^4\Bar{F}\right]+\frac{1}{\zeta^5}\left[F\cdot (\nabla^2 F)^2-\frac{1}{2} F^2\cdot \nabla^4 F\right] \right\}
\end{align}
The action has projective covariant measures, and its integration contour on $\mathbb{CP}^1$ ensures invariance under $SU(2)$ transformation. The parts with $F$ and $\bar{F}$ transform to each other through Hermitian antipodal map $\zeta\rightarrow-1/\zeta$.\\

\textbf{Consistency with the Abelian form.} We can check whether the above received non-Abelian action would reduce to the Abelian action obtained in (\ref{5dAbCS}). First, we notice that for the Abelian case, $F|_{Ab}=\oint d\zeta V(\zeta)=v_{-1}$ and $\bar{F}|_{Ab}=\oint d\zeta \frac{V(\zeta)}{\zeta^2}=v_1$. Here, we recall that the vector multiplet is tropical: $V(\zeta)=\sum_{n=-\infty}^{\infty}v_n\zeta^n$; gauge parameters $\Lambda$ and $\bar{\Lambda}$ are arctic/antarctic multiplets: $\Lambda=\sum_{n=0}^\infty\lambda_n\zeta^n$, $\bar{\Lambda}=\sum_{n=0}^\infty\bar{\lambda}_{-n}\zeta^{-n}$ of which, the lowest two and the highest two components are constrained, respectively. The constraints are: \(D_1\lambda_0=D_1^\alpha D_{1\alpha}\lambda_1=0 ,\; D_2\bar{\lambda}_0=D_2^\alpha D_{2\alpha}\bar{\lambda}_{-1}=0\). The infinitesimal Abelian gauge transformation of the vector multiplet $V$ is:
\[
\delta V=i(\bar{\Lambda}-\Lambda) \implies \delta v_0=i(\bar{\lambda}_0-\lambda_0),\; \delta v_n=-\lambda_n,\; \delta v_{-n}=\bar{\lambda}_n
\]
Then, we can set all $v_n=0, \; \forall n\neq -1,0,1$, making $V$ a $\mathcal{O}(2)$ multiplet and then further gauging away all the unconstrained $v_1$ and $v_{-1}$, leaving only $D_1D_1v_1$ and $D_2D_2 v_{-1}$ pieces, setting the vector multiplet in the Lindstr\"om-Ro\v{c}ek gauge \cite{Lindstrom:1989ne}. As we have seen before, these pieces are the Abelian field strength. $D_2D_2v_{-1}=D_1D_1v_1=\mathcal{W}^{Ab}$. Using the gauge, further simplifications are made in the Abelian action. In $\zeta$ components the action (\ref{5dAbCS}) becomes:
\begin{align}
S_{Ab}\sim\int d^5 x d^8\theta \oint d\zeta_0 V_0 A_{\zeta_0}\nabla_0^2 A_{\zeta_0}=\frac{1}{2}\int d^5 x d^8\theta v_{-1}v_1 (D_1D_1v_1+D_2D_2 v_{-1})
\end{align}
The full superspace measure is: $d^8\theta=(D_1D_1)^2(D_2D_2)^2$. Using the projectivity of the vector multiplet and the Lindstr\"om-Ro\v{c}ek gauge, we find: $D_1v_{-1}=D_2v_{-2}=0,\;D_2v_1=D_1v_2=0$. This allows us to write:
\[
(D_1D_1)[v_{-1}v_1D_2D_2v_{-1}]=v_{-1}(D_2D_2v_{-1})^2\;;\;\;(D_2D_2)[v_{-1}v_1D_1D_1v_{1}]=v_{1}(D_1D_1v_{1})^2\;,
\]
which are the first terms of $F$ and $\bar{F}$ parts, respectively in eq.(\ref{NonAb_CS}). Furthermore, for the Abelian case, the second terms of both the $F$ and $\bar{F}$ parts will be zero: \(v_{-1}^2(D_2D_2)^2v_{-1}=v_{-1}^2(D_2D_2)D_1D_1v_{1}=0\) and \(v_1^2(D_1D_1)^2v_1=0\).

Thus, we have proven that the non-Abelian action in (\ref{NonAb_CS}) will reduce to the form equivalent to equation (\ref{5dAbCS}) in the Abelian limit.

\subsection{Discussion}
In this example (however it is not a mere example, but it has been an unsolved problem for a long time) we saw how useful it is to use the asymmetric splitting techniques of the projective superspace to decompose the prepotential function $e^V$ and subsequently express the field strength in terms of the $e^{\check{U}}$ (or $e^{\hat{U}}$) projections and the $\zeta$-independent term $D^{\hat{\alpha}}_1D_{\hat{\alpha}1}\bar{F}$ (respectively $D^{\hat{\alpha}}_2D_{\hat{\alpha}2}F$). The advantage of doing this was that it is now easier to separate terms, and the contour-integral eliminated unwanted terms and left us with much simpler expressions. Furthermore, analysis on the ordinary covariant derivatives in 5-dimensions and applying partial integrations on them, we can finally pull out the variation and find the long-awaited non-Abelian Chern-Simons action in terms of functionals $F$ (\ref{F}) and $\bar{F}$ (\ref{barF}) that depend on unconstrained prepotential superfields $X_i=e^{V_i}-1$. Our final result is written in terms of the $\zeta$-independent functionals, these are themselves highly nontrivial functionals of the gauge prepotential $V(\zeta_i)$ built through contour-integral projections. Since direct component expansions of F and $\bar{F}$ are intractable, we only demonstrated the consistency of the non-Abelian action after reducing to the Abelian limit with the earlier obtained Abelian action in \ref{5dAbCS}. It is only thanks to the contour integrating and splittig techniques on the physical prepotential $e^V$ and the field strenth $\mathcal{W}$, we have been able to integrate this variational form of action. Unfortunately, after obtaining the integrated lagrangian, it becomes very hard to restore the $\zeta$-dependent multiplets such as field strengths and vector multiplets. However, they are related to the field strengths and the posivitive/negative projections of the prepotential exponential through the eq.(\ref{relation:W:F}).

%% file: Acknowledgements.tex
\section*{Acknowledgements}\label{sec:Acknowledge}
The author would like to thank Rikard von Unge for fruitful discussions and Silvia Penati and Sergei Kuzenko for reading the earlier version of this article. Also, I am grateful for their kind hospitality to INFN, Padova. 

%% file: sections/Appendix.tex
\section{Nonsymmetric splitting of the prepotential}
Typically, one uses the arctic and antarctic prescription \cite{Davgadorj:2017ezp} to describe the field strength, where the prepotential $e^V$ is decomposed into the parts containing the negative and positive  powers of $\zeta$, respectively, $e^{\bar{U}}$ and $e^U$. Those two parts are projectively conjugate to each other and the half of the $\zeta$-independent terms sit in the $e^{\bar{U}}$ and the another half in the $e^U$ part, therefore symmetric. Then we could consider what if we let the zero-mode ($\zeta$ independent) terms to be contained in either the negative or positive $\zeta$-power part. \\
In the symmetric splitting:
\begin{equation}
    e^V=e^{\bar{U}}e^U
\end{equation}
if we expand $e^U$ and $e^{\bar{U}}$ in powers of $X=e^V-1$ as a sum:
\begin{align}
    & e^U=1+Y^{(1)}+Y^{(2)}+\dots \\
    & e^{\bar{U}}=1+\bar{Y}^{(1)}+\bar{Y}^{(2)}+\dots
\end{align}
we get an infinite set of equations that can be solved recursively:
\begin{align}
    \bar{Y}^{(1)}+Y^{(1)}&=X \nn \\
    \bar{Y}^{(2)}+Y^{(2)}&=-\bar{Y}^{(1)}Y^{(1)} \nn \\
    & \vdots \nn \\
    \bar{Y}^{(n)}+Y^{(n)}&=-\bar{Y}^{(n-1)}Y^{(1)}-\dots-\bar{Y}^{(1)}Y^{(n-1)}
\end{align}
In each stage of the solution, we need to project the expression on the right hand side on positive or negative powers of $\zeta$ with a symmetric splitting of the zero-mode terms so that the $\bar{Y}^{(n)}$ term is the conjugate of the $Y^{(n)}$. As an example we can write down the first two terms and see that this splitting is complicated.
\begin{align}\label{sym.projections}
    &Y^{(1)}=X^+-\frac{1}{2}x_0\;\;,\;\;\bar{Y}^{(1)}=X^-+\frac{1}{2}x_0 \\
    &Y^{(2)}=-[X^-X^+]^{+}+\frac{1}{2}[X^-X^+]^0-\frac{1}{2}x_0X^++\frac{3}{8}x_0^2 \\
    &\bar{Y}^{(2)}=-[X^-X^+]^{-}-\frac{1}{2}[X^-X^+]^0+\frac{1}{2}X^-x_0-\frac{1}{8}x_0^2
\end{align}
here, $X^+$ contains positive powers plus zero-modes of $\zeta$, $X^-$ contains only negative powers.\\
So it turns out that the projections by $\zeta$-integrals and the $\epsilon$-prescription naturally favors nonsymmetric projections.
\begin{align}
    &\oint d\zeta_1\frac{X_1}{\zeta_{10}}=\sum^\infty_{n=0}x_n\zeta_0^n \nn \\
    &\oint d\zeta_1\frac{X_1}{\zeta_{01}}=\sum^{-1}_{n=-\infty}x_n\zeta_0^n
\end{align}
This inspires to split the prepotential as: $e^V=e^{\check{\bar
U}}e^{\hat{U}}$, where $e^{\hat{U}}$ contains positive plus all zero-mode terms (at each powers of X). Then the above recursion relations can be written in terms of:
\begin{align}
    &e^{\hat{U}}=1+\hat{Y}^{(1)}+\hat{Y}^{(2)}+\dots \\
    & e^{\check{\bar{U}}}=1+\check{\bar{Y}}^{(1)}+\check{\bar{Y}}^{(2)}+\dots
\end{align}
as sums of only negative terms and positive terms containing zero-mode terms.
\begin{align}
    \check{\bar{Y}}^{(1)}+\hat{Y}^{(1)}&=X \nn \\
    \check{\bar{Y}}^{(2)}+\hat{Y}^{(2)}&=-\check{\bar{Y}}^{(1)}\hat{Y}^{(1)} \nn \\
    & \vdots \nn \\
    \check{\bar{Y}}^{(n)}+\hat{Y}^{(n)}&=-\check{\bar{Y}}^{(n-1)}\hat{Y}^{(1)}-\dots-\check{\bar{Y}}^{(1)}\hat{Y}^{(n-1)}
\end{align}
This allows us to write compactly using contour integral projections as:
\begin{align}
    e^{\hat{U}}=& 1+\sum^\infty_{n=1}(-1)^{n+1}\oint d\zeta_1\dots d\zeta_n\frac{X_1\dots X_n}{\zeta_{21}\dots \zeta_{n,n-1}}\frac{1}{\zeta_{n0}} \\
    e^{\check{\bar{U}}}=& 1+\sum^\infty_{n=1}(-1)^{n+1}\oint d\zeta_1\dots d\zeta_n \frac{1}{\zeta_{01}}\frac{X_1\dots X_n}{\zeta_{21}\dots \zeta_{n,n-1}}
\end{align}
In the language of simple projections (\ref{sym.projections}) the above can be written:
\begin{align}
    &e^{\hat{U}}=1+X^+-[X^-X]^{+}+\left[[X^-X]^-X\right]^{+}-\dots \nn \\
    &e^{\check{\bar{U}}}=1+X^- -[XX^+]^- + \left[X[XX^+]^+\right]^- -\dots
\end{align}
One could also be interested in putting the zero-mode terms together with the negative power terms as in:
\begin{align}
    &\oint d\zeta_1\frac{X_1}{\zeta_{10}}\frac{\zeta_0}{\zeta_1}=\sum^\infty_{n=1}x_n\zeta_0^n \nn \\
    &\oint d\zeta_1\frac{X_1}{\zeta_{01}}\frac{\zeta_0}{\zeta_1}=\sum^{0}_{n=-\infty}x_n\zeta_0^n
\end{align}
This kind of projection suggests to split the prepotential as $e^V=e^{\hat{\bar
U}}e^{\check{U}}$, where $e^{\check{U}}$ does not contain zero-mode. Doing the same recursion as above and reminding yourself the additional $\frac{\zeta_0}{\zeta_1}$ projection, which will shift terms like this and remove the zero-mode terms:
\begin{equation}
    \frac{1}{\zeta_{n,0}}\frac{\zeta_0}{\zeta_1}=\frac{1}{\zeta_n}\sum_{k=0}^\infty \left(\frac{\zeta_0}{\zeta_n}\right)^k\frac{\zeta_0}{\zeta_1}=\frac{1}{\zeta_n}\sum_{k=1}^\infty \left(\frac{\zeta_0}{\zeta_n}\right)^k\frac{\zeta_n}{\zeta_1}
\end{equation}
The contour integral projections are the following.
\begin{align}
    e^{\check{U}}=& 1+\sum^\infty_{n=1}(-1)^{n+1}\oint d\zeta_1\dots d\zeta_n\frac{\zeta_0}{\zeta_1}\frac{X_1\dots X_n}{\zeta_{21}\dots \zeta_{n,n-1}}\frac{1}{\zeta_{n0}} \\
    e^{\hat{\bar{U}}}=& 1+\sum^\infty_{n=1}(-1)^{n+1}\oint d\zeta_1\dots d\zeta_n \frac{1}{\zeta_{01}}\frac{X_1\dots X_n}{\zeta_{21}\dots \zeta_{n,n-1}}\frac{\zeta_0}{\zeta_n}
\end{align}
In the language of simple projections (\ref{sym.projections}) the above are:
\begin{align}
    &e^{\check{U}}-1=\left[\frac{X}{\zeta}\right]^{+}\zeta-\left[\left[\frac{X}{\zeta}\right]^{-}X\right]^{+}\zeta+\left[\left[\left[\frac{X}{\zeta}\right]^{-}X\right]^{-}X\right]^{+}\zeta-\dots \nn \\
    &=\left(X^+ -x_0-[X^-X^+]^+ -x_0X^+ +\left[[X^-X]^-X^+\right]^+ +x_0[X^-X^+]^+ \dots\right)\zeta \nn \\
    &e^{\hat{\bar{U}}}-1=\left[\frac{X}{\zeta}\right]^{-}\zeta-\left[X\left[\frac{X}{\zeta}\right]^{+}\right]^{-}\zeta+\left[X\left[X\left[\frac{X}{\zeta}\right]^{+}\right]^{+}\right]^{-}\zeta-\dots \nn \\
    &=\left(X^-+x_0 -[X^-X^+]^- +X^-x_0 +\left[X^-[XX^+]^+\right]^- -[X^-X^+]^-x_0\dots\right)\zeta
\end{align}
For completeness of the description and for the practical uses later we give also the inverses.
\begin{align}
    e^{-\hat{U}}=& 1+\sum^\infty_{n=1}(-1)^{n}\oint d\zeta_1\dots d\zeta_n\frac{1}{\zeta_{10}}\frac{X_1\dots X_n}{\zeta_{21}\dots \zeta_{n,n-1}} \\
    e^{-\check{\bar{U}}}=& 1+\sum^\infty_{n=1}(-1)^{n}\oint d\zeta_1\dots d\zeta_n \frac{X_1\dots X_n}{\zeta_{21}\dots \zeta_{n,n-1}}\frac{1}{\zeta_{0n}} \\
    e^{-\check{U}}=& 1+\sum^\infty_{n=1}(-1)^{n}\oint d\zeta_1\dots d\zeta_n\frac{1}{\zeta_{10}}\frac{X_1\dots X_n}{\zeta_{21}\dots \zeta_{n,n-1}}\frac{\zeta_0}{\zeta_{n}} \\
    e^{-\hat{\bar{U}}}=& 1+\sum^\infty_{n=1}(-1)^{n}\oint d\zeta_1\dots d\zeta_n \frac{\zeta_0}{\zeta_{1}}\frac{X_1\dots X_n}{\zeta_{21}\dots \zeta_{n,n-1}}\frac{1}{\zeta_{0n}}
\end{align}
To be clear let us write these different splittings together in one place.
\begin{equation} \label{splittings}
    e^V=e^{\bar{U}}e^U=e^{\check{\bar{U}}}e^{\hat{U}}=e^{\hat{\bar{U}}}e^{\check{U}}
\end{equation}
They differ by where the $\zeta$-independent terms sit, therefore one can relate those splittings through a $\zeta$-independent object as in $e^{\hat{U}}=e^P e^U$. From (\ref{splittings}) we can write this $e^P$ and its conjugate $e^{\bar{P}}$ equivalently:
\begin{align}
    &e^P=e^{\hat{U}}e^{-U}=e^{-\check{\bar{U}}}e^{\bar{U}} \\
    &e^{\bar{P}}=e^{U}e^{-\check{U}} \\
    &e^Pe^{\bar{P}}=e^{\hat{U}}e^{-\check{U}}
\end{align}
Employing again recursions we arrive at the concise expression for the difference between the zero-mode containing and the no-zero-mode bridges.
\begin{equation}\label{eP}
    e^Pe^{\bar{P}}=1+\sum^\infty_{n=1}(-1)^{n+1}\oint d\zeta_1\dots d\zeta_n\frac{X_1\dots X_n}{\zeta_{21}\dots \zeta_{n,n-1}}\frac{1}{\zeta_n}
\end{equation}
its inverse is naturally as expected
\begin{equation} \label{inveP}
    e^{-\bar{P}}e^{-P}=1+\sum^\infty_{n=1}(-1)^{n+1}\oint d\zeta_1\dots d\zeta_n\frac{1}{\zeta_1}\frac{X_1\dots X_n}{\zeta_{21}\dots \zeta_{n,n-1}}
\end{equation}
Because this $e^P$ is $\zeta$-independent, consequently the gauge connection $A_\zeta$ does not depend on which splitting we use.
\begin{equation}
    A_\zeta=e^{-U}(\partial_\zeta e^U)=e^{-\hat{U}}(\partial_\zeta e^{\hat{U}})=e^{-\check{U}}(\partial_\zeta e^{\check{U}})
\end{equation}
After finding out the explicit form of the connection $A_\zeta$ we will write down the field strength in terms of the nonsymmetric bridges $e^{\hat{U}}$ and $e^{\check{U}}$.

\section{Field strength in terms of nonsymmetric splittings}\label{W-nonsym}

Now that we are equipped with the explicit contour integral form of the gauge connection and these new projecting devices (nonsymmetric splitting), we would like to write down the polar field strength $i\mathcal{W}=\Bar{\nabla}^2 A$ in terms of this new language. In order to do that, first, we need to express it in terms of $\mathcal{N}=1$ derivatives.
\begin{align}
    \Bar{\nabla}_0^2 A_0=&\Bar{\nabla}_0^2\sum_{n=1}^\infty\frac{(-1)^{n+1}X_1\dots X_n}{\zeta_{10}\zeta_{21}\dots\zeta_{n,n-1}\zeta_{n0}}= \nn \\
    =&\sum_{n=1}^\infty (-1)^{n+1}\frac{1}{2}\sum_{k=1}^n\sum_{i=1}^n\frac{X_1\dots\Bar{\nabla}_0^\alpha X_k\dots\Bar{\nabla}_{0\alpha}X_i\dots X_n}{\zeta_{10}\zeta_{21}\dots\zeta_{n,n-1}\zeta_{n0}} 
\end{align}
Projectivity of fields gives the relations:
\begin{align}
    &\Bar{\nabla}_0 X_1=(\zeta_1-\zeta_0)\Bar{D}X_1 \nonumber\\
    &\Bar{\nabla}_0^2X_1=(\zeta_1-\zeta_0)^2\Bar{D}^2X_1
\end{align}
Also, when one of $\nabla_0$'s hits the $X_k$, we decompose the $\zeta_k-\zeta_0$ factor by shifting left or right depending on which side the given $\nabla$ sits.
\begin{align}
    &\zeta_k-\zeta_0=(\zeta_k-\zeta_{k-1})+(\zeta_{k-1}-\zeta_{k-2})+\dots(\zeta_2-\zeta_{1})+(\zeta_1-\zeta_0) \nonumber\\
    &\zeta_i-\zeta_0=(\zeta_i-\zeta_{i+1})+(\zeta_{i+1}-\zeta_{i+2})+\dots(\zeta_{n-1}-\zeta_{n})+(\zeta_n-\zeta_0)
\end{align}
Each of these factors cancel one factor from the denominator, and this empty spot in the string of factors in the denominator indicates where we separate the part between the $\bar{D}^2$ free left (right) side and the middle part with the $\bar{D}^2$ derivatives.
\begin{align}
    \Bar{\nabla}_0^2 A_0=\sum_{n=1}^\infty\sum_{a=0}^{b-1}\sum_{b=1}^{n}\oint &\frac{(-1)^a X_1\dots X_a}{\zeta_{10}\dots\zeta_{a,a-1}} \cdot\frac{(-1)^{b-(a+1)}\Bar{D}^2(X_{a+1}\dots X_{b})}{\zeta_{a+2,a+1}\dots\zeta_{b,b-1}}\cdot \nn \\
    &\cdot\frac{(-1)^{n+1-(b+1)}X_{b+1}\dots X_n}{\zeta_{b+2,b+1}\dots\zeta_{n0}}
\end{align} 
On the left and right side of the $\Bar{D}^2$ we recognize the $e^{-\hat{U}}$ and $e^{\hat{U}}$ sums, therefore:
\begin{align}\label{def:F}
    \Bar{\nabla}_0^2\mathcal{A}_0=e^{-\hat{U}_0}\Bar{D}^2\sum_{b=1}^\infty\oint d\zeta_1\dots d\zeta_b\frac{(-1)^{b-1}X_1\dots X_b}{\zeta_{21}\dots\zeta_{b,b-1}}e^{\hat{U}_0}\equiv e^{-\hat{U}_0}\Bar{D}^2 F e^{\hat{U}_0}\;,
\end{align}
where we defined a new $\zeta$ independent sum $F$:
\begin{align} \label{F}
   F=\sum^\infty_{n=1}(-1)^{n+1}\oint d\zeta_1\dots d\zeta_n \frac{X_1\dots X_n}{\zeta_{21}\dots \zeta_{n,n-1}} 
\end{align}

We could have equivalently done the above in terms of $\bar{Q}$ derivatives.
\begin{align}
    &\Bar{\nabla}_0 X_1=\frac{\zeta_1-\zeta_0}{\zeta_1}\Bar{Q}X_1 \nonumber\\
    &\Bar{\nabla}_0^2X_1=\frac{(\zeta_1-\zeta_0)^2}{\zeta_1^2}\Bar{Q}^2X_1
\end{align}
By doing similar decompositions and separating terms, we will receive another expression for the field strength. 
\begin{align}
    i\mathcal{W}=e^{-\check{U}}\Bar{Q}^2 \Bar{F} e^{\check{U}}
\end{align}
here the $\zeta$-independent term $\bar{F}$ is found to be:
\begin{align}\label{barF}
    \Bar{F}=\sum^\infty_{n=1}(-1)^{n+1}\oint d\zeta_1\dots d\zeta_n \frac{1}{\zeta_1} \frac{X_1\dots X_n}{\zeta_{21}\dots \zeta_{n,n-1}}\frac{1}{\zeta_n}
\end{align}
Summarizing them:
\begin{align} \label{W_nonsymm}
    &i\mathcal{W}=e^{-\hat{U}}\Bar{D}^2 F e^{\hat{U}}=e^{-\check{U}}\Bar{Q}^2 \Bar{F} e^{\check{U}}=e^{-U}e^{-P}\Bar{D}^2 F e^P e^U=e^{-U}e^{\bar{P}}\Bar{Q}^2 \Bar{F} e^{-\bar{P}} e^U \nn \\
    &i\Bar{\mathcal{W}}=e^{-\hat{U}}Q^2 F e^{\hat{U}}=e^{-\check{U}}D^2 \Bar{F} e^{\check{U}}=e^{-U}e^{-P}Q^2 F e^P e^U=e^{-U}e^{\bar{P}}D^2 \Bar{F} e^{-\bar{P}} e^U
\end{align}

Then from (\ref{W_nonsymm}) it is clear that the field strength in the vector representation can be written as:
\begin{align}
    \mathbb{W}=&e^{-P}\Bar{D}^2 F e^{P}=e^{\bar{P}}\Bar{Q}^2\Bar{F}e^{-\bar{P}} \nn\\
    \Bar{\mathbb{W}}=&e^{\bar{P}}D^2\Bar{F}e^{-\bar{P}}=e^{-P}Q^2 F e^{P}
\end{align}